\documentclass[a4paper]{article} 

\usepackage[draft]{minted}
\setminted{breaklines}
\newminted{haskell}{}
\usepackage{geometry}
\usepackage{amsmath}
\usepackage{amssymb}
\usepackage{graphicx}
\usepackage{algpseudocode}

\newlength{\extraplusheight}
\newlength{\extrapluswidth}
\newcommand\mdoubleplus{\mathbin{+\mkern-10mu+}}

\newtheorem{theorem}{Theorem}[section]

\newtheorem{lemma}[theorem]{Lemma}
\newtheorem{example}[theorem]{Example}

\newtheorem{definition}[theorem]{Definition}

\newcommand{\msf}[1]{\mbox{{\sf #1}}}

\newcommand{\Set}{\msf{Set}}
\newcommand{\Perm}{\msf{Perm}}
\newcommand{\Nat}{\msf{Nat}}

\newcommand{\Fin}{\msf{Fin}}

\makeatletter
\def\PYG@reset{\let\PYG@it=\relax \let\PYG@bf=\relax%
    \let\PYG@ul=\relax \let\PYG@tc=\relax%
    \let\PYG@bc=\relax \let\PYG@ff=\relax}
\def\PYG@tok#1{\csname PYG@tok@#1\endcsname}
\def\PYG@toks#1+{\ifx\relax#1\empty\else%
    \PYG@tok{#1}\expandafter\PYG@toks\fi}
\def\PYG@do#1{\PYG@bc{\PYG@tc{\PYG@ul{%
    \PYG@it{\PYG@bf{\PYG@ff{#1}}}}}}}
\def\PYG#1#2{\PYG@reset\PYG@toks#1+\relax+\PYG@do{#2}}

\@namedef{PYG@tok@w}{\def\PYG@tc##1{\textcolor[rgb]{0.73,0.73,0.73}{##1}}}
\@namedef{PYG@tok@c}{\let\PYG@it=\textit\def\PYG@tc##1{\textcolor[rgb]{0.24,0.48,0.48}{##1}}}
\@namedef{PYG@tok@cp}{\def\PYG@tc##1{\textcolor[rgb]{0.61,0.40,0.00}{##1}}}
\@namedef{PYG@tok@k}{\let\PYG@bf=\textbf\def\PYG@tc##1{\textcolor[rgb]{0.00,0.50,0.00}{##1}}}
\@namedef{PYG@tok@kp}{\def\PYG@tc##1{\textcolor[rgb]{0.00,0.50,0.00}{##1}}}
\@namedef{PYG@tok@kt}{\def\PYG@tc##1{\textcolor[rgb]{0.69,0.00,0.25}{##1}}}
\@namedef{PYG@tok@o}{\def\PYG@tc##1{\textcolor[rgb]{0.40,0.40,0.40}{##1}}}
\@namedef{PYG@tok@ow}{\let\PYG@bf=\textbf\def\PYG@tc##1{\textcolor[rgb]{0.67,0.13,1.00}{##1}}}
\@namedef{PYG@tok@nb}{\def\PYG@tc##1{\textcolor[rgb]{0.00,0.50,0.00}{##1}}}
\@namedef{PYG@tok@nf}{\def\PYG@tc##1{\textcolor[rgb]{0.00,0.00,1.00}{##1}}}
\@namedef{PYG@tok@nc}{\let\PYG@bf=\textbf\def\PYG@tc##1{\textcolor[rgb]{0.00,0.00,1.00}{##1}}}
\@namedef{PYG@tok@nn}{\let\PYG@bf=\textbf\def\PYG@tc##1{\textcolor[rgb]{0.00,0.00,1.00}{##1}}}
\@namedef{PYG@tok@ne}{\let\PYG@bf=\textbf\def\PYG@tc##1{\textcolor[rgb]{0.80,0.25,0.22}{##1}}}
\@namedef{PYG@tok@nv}{\def\PYG@tc##1{\textcolor[rgb]{0.10,0.09,0.49}{##1}}}
\@namedef{PYG@tok@no}{\def\PYG@tc##1{\textcolor[rgb]{0.53,0.00,0.00}{##1}}}
\@namedef{PYG@tok@nl}{\def\PYG@tc##1{\textcolor[rgb]{0.46,0.46,0.00}{##1}}}
\@namedef{PYG@tok@ni}{\let\PYG@bf=\textbf\def\PYG@tc##1{\textcolor[rgb]{0.44,0.44,0.44}{##1}}}
\@namedef{PYG@tok@na}{\def\PYG@tc##1{\textcolor[rgb]{0.41,0.47,0.13}{##1}}}
\@namedef{PYG@tok@nt}{\let\PYG@bf=\textbf\def\PYG@tc##1{\textcolor[rgb]{0.00,0.50,0.00}{##1}}}
\@namedef{PYG@tok@nd}{\def\PYG@tc##1{\textcolor[rgb]{0.67,0.13,1.00}{##1}}}
\@namedef{PYG@tok@s}{\def\PYG@tc##1{\textcolor[rgb]{0.73,0.13,0.13}{##1}}}
\@namedef{PYG@tok@sd}{\let\PYG@it=\textit\def\PYG@tc##1{\textcolor[rgb]{0.73,0.13,0.13}{##1}}}
\@namedef{PYG@tok@si}{\let\PYG@bf=\textbf\def\PYG@tc##1{\textcolor[rgb]{0.64,0.35,0.47}{##1}}}
\@namedef{PYG@tok@se}{\let\PYG@bf=\textbf\def\PYG@tc##1{\textcolor[rgb]{0.67,0.36,0.12}{##1}}}
\@namedef{PYG@tok@sr}{\def\PYG@tc##1{\textcolor[rgb]{0.64,0.35,0.47}{##1}}}
\@namedef{PYG@tok@ss}{\def\PYG@tc##1{\textcolor[rgb]{0.10,0.09,0.49}{##1}}}
\@namedef{PYG@tok@sx}{\def\PYG@tc##1{\textcolor[rgb]{0.00,0.50,0.00}{##1}}}
\@namedef{PYG@tok@m}{\def\PYG@tc##1{\textcolor[rgb]{0.40,0.40,0.40}{##1}}}
\@namedef{PYG@tok@gh}{\let\PYG@bf=\textbf\def\PYG@tc##1{\textcolor[rgb]{0.00,0.00,0.50}{##1}}}
\@namedef{PYG@tok@gu}{\let\PYG@bf=\textbf\def\PYG@tc##1{\textcolor[rgb]{0.50,0.00,0.50}{##1}}}
\@namedef{PYG@tok@gd}{\def\PYG@tc##1{\textcolor[rgb]{0.63,0.00,0.00}{##1}}}
\@namedef{PYG@tok@gi}{\def\PYG@tc##1{\textcolor[rgb]{0.00,0.52,0.00}{##1}}}
\@namedef{PYG@tok@gr}{\def\PYG@tc##1{\textcolor[rgb]{0.89,0.00,0.00}{##1}}}
\@namedef{PYG@tok@ge}{\let\PYG@it=\textit}
\@namedef{PYG@tok@gs}{\let\PYG@bf=\textbf}
\@namedef{PYG@tok@gp}{\let\PYG@bf=\textbf\def\PYG@tc##1{\textcolor[rgb]{0.00,0.00,0.50}{##1}}}
\@namedef{PYG@tok@go}{\def\PYG@tc##1{\textcolor[rgb]{0.44,0.44,0.44}{##1}}}
\@namedef{PYG@tok@gt}{\def\PYG@tc##1{\textcolor[rgb]{0.00,0.27,0.87}{##1}}}
\@namedef{PYG@tok@err}{\def\PYG@bc##1{{\setlength{\fboxsep}{\string -\fboxrule}\fcolorbox[rgb]{1.00,0.00,0.00}{1,1,1}{\strut ##1}}}}
\@namedef{PYG@tok@kc}{\let\PYG@bf=\textbf\def\PYG@tc##1{\textcolor[rgb]{0.00,0.50,0.00}{##1}}}
\@namedef{PYG@tok@kd}{\let\PYG@bf=\textbf\def\PYG@tc##1{\textcolor[rgb]{0.00,0.50,0.00}{##1}}}
\@namedef{PYG@tok@kn}{\let\PYG@bf=\textbf\def\PYG@tc##1{\textcolor[rgb]{0.00,0.50,0.00}{##1}}}
\@namedef{PYG@tok@kr}{\let\PYG@bf=\textbf\def\PYG@tc##1{\textcolor[rgb]{0.00,0.50,0.00}{##1}}}
\@namedef{PYG@tok@bp}{\def\PYG@tc##1{\textcolor[rgb]{0.00,0.50,0.00}{##1}}}
\@namedef{PYG@tok@fm}{\def\PYG@tc##1{\textcolor[rgb]{0.00,0.00,1.00}{##1}}}
\@namedef{PYG@tok@vc}{\def\PYG@tc##1{\textcolor[rgb]{0.10,0.09,0.49}{##1}}}
\@namedef{PYG@tok@vg}{\def\PYG@tc##1{\textcolor[rgb]{0.10,0.09,0.49}{##1}}}
\@namedef{PYG@tok@vi}{\def\PYG@tc##1{\textcolor[rgb]{0.10,0.09,0.49}{##1}}}
\@namedef{PYG@tok@vm}{\def\PYG@tc##1{\textcolor[rgb]{0.10,0.09,0.49}{##1}}}
\@namedef{PYG@tok@sa}{\def\PYG@tc##1{\textcolor[rgb]{0.73,0.13,0.13}{##1}}}
\@namedef{PYG@tok@sb}{\def\PYG@tc##1{\textcolor[rgb]{0.73,0.13,0.13}{##1}}}
\@namedef{PYG@tok@sc}{\def\PYG@tc##1{\textcolor[rgb]{0.73,0.13,0.13}{##1}}}
\@namedef{PYG@tok@dl}{\def\PYG@tc##1{\textcolor[rgb]{0.73,0.13,0.13}{##1}}}
\@namedef{PYG@tok@s2}{\def\PYG@tc##1{\textcolor[rgb]{0.73,0.13,0.13}{##1}}}
\@namedef{PYG@tok@sh}{\def\PYG@tc##1{\textcolor[rgb]{0.73,0.13,0.13}{##1}}}
\@namedef{PYG@tok@s1}{\def\PYG@tc##1{\textcolor[rgb]{0.73,0.13,0.13}{##1}}}
\@namedef{PYG@tok@mb}{\def\PYG@tc##1{\textcolor[rgb]{0.40,0.40,0.40}{##1}}}
\@namedef{PYG@tok@mf}{\def\PYG@tc##1{\textcolor[rgb]{0.40,0.40,0.40}{##1}}}
\@namedef{PYG@tok@mh}{\def\PYG@tc##1{\textcolor[rgb]{0.40,0.40,0.40}{##1}}}
\@namedef{PYG@tok@mi}{\def\PYG@tc##1{\textcolor[rgb]{0.40,0.40,0.40}{##1}}}
\@namedef{PYG@tok@il}{\def\PYG@tc##1{\textcolor[rgb]{0.40,0.40,0.40}{##1}}}
\@namedef{PYG@tok@mo}{\def\PYG@tc##1{\textcolor[rgb]{0.40,0.40,0.40}{##1}}}
\@namedef{PYG@tok@ch}{\let\PYG@it=\textit\def\PYG@tc##1{\textcolor[rgb]{0.24,0.48,0.48}{##1}}}
\@namedef{PYG@tok@cm}{\let\PYG@it=\textit\def\PYG@tc##1{\textcolor[rgb]{0.24,0.48,0.48}{##1}}}
\@namedef{PYG@tok@cpf}{\let\PYG@it=\textit\def\PYG@tc##1{\textcolor[rgb]{0.24,0.48,0.48}{##1}}}
\@namedef{PYG@tok@c1}{\let\PYG@it=\textit\def\PYG@tc##1{\textcolor[rgb]{0.24,0.48,0.48}{##1}}}
\@namedef{PYG@tok@cs}{\let\PYG@it=\textit\def\PYG@tc##1{\textcolor[rgb]{0.24,0.48,0.48}{##1}}}

\def\PYGZgt{\char`\>}

\def\PYGZhy{\char`\-}

\makeatother

\title{{\bf Filter Equivariant Functions}\\ A symmetric account of length-general extrapolation on lists
}

\author{Owen Lewis$^{2*}$, Neil Ghani$^{3,4*}$, Andrew Dudzik$^1$, Christos Perivolaropoulos$^1$,\\ Razvan Pascanu$^1$ and Petar Veli\v{c}kovi\'{c}$^1$\\ \\
$^1$Google DeepMind \qquad $^2$Goodfire AI \qquad $^3$Kodamai\\ $^4$University of Strathclyde \qquad  $^*$Work done at Google DeepMind}

\date{}

\begin{document}
\maketitle

\begin{abstract}
\noindent What should a function that \emph{extrapolates} beyond known input/output examples look like? This is a tricky question to answer in general, as any function matching the outputs on those examples can in principle be a correct extrapolant. We argue that a ``good'' extrapolant should follow certain kinds of \emph{rules}, and here we study a particularly appealing criterion for rule-following in list functions: that the function should behave predictably even when certain elements are \emph{removed}. In functional programming, a standard way to express such removal operations is by using a \msf{filter} function. Accordingly, our paper introduces a new semantic class of functions -- the \msf{filter} equivariant functions. We show that this class contains interesting examples, prove some basic theorems about it, and relate it to the well-known class of \msf{map} equivariant functions. We also present a geometric account  of \msf{filter} equivariants, showing how they correspond naturally to certain simplicial structures. Our highlight result is the \emph{amalgamation} algorithm, which constructs any \msf{filter}-equivariant function's output by first studying how it behaves on sublists of the input, in a way that extrapolates perfectly.
% Our highlight result is the \emph{amalgamation} algorithm, which can construct the output of a filter equivariant function by studying how it behaves on the sublists. We show that this algorithm, due to the filter equivariance property, extrapolates perfectly. 

% show it contains interesting examples, and prove some basic theorems about this class. We also relate this class to the well . We show this class of functions also supports an {\em amalgamation} property meaning its behaviour on large inputs can be computed
% from its behaviour on small inputs, crucial in machine learning where one cannot train on arbitrary large examples.
\end{abstract}

\section{Introduction}

We want to mathematically characterise functions with certain kinds of \emph{rule-following} behaviour. The nature of rule-following is a longstanding philosophical riddle: given some pattern of behavior, what exactly does it mean to extend or \emph{extrapolate} it consistently? While intuitively clear, the notion of extrapolation has been shown to be difficult or impossible to define in general. Exemplifying a common line of argumentation, Saul Kripke, following Wittgenstein, devises a pathological mathematical operation, ``$\mathrm{quus}(x, y)$'', which operates like ordinary addition if both $x$ and $y$ are less than $57$, and returns $5$ otherwise \cite{Kripke1982-KRIWOR}. Kripke imagines a student who, having never before added numbers greater than 57, answers 125 when asked for 68 + 57. How can they be so sure, asks Kripke? After all, all of their previous small-number behavior was consistent with enacting either $+$ or quus; what principle could justify one extrapolation over the other?

After considering and rejecting several alternatives, Kripke reaches the conclusion that there can be no such principle capable of surviving skeptical attack. Put differently, \emph{any} function fitting the observed examples correctly can be a valid extrapolant. 

% To make more sense of which functions make more sense than others, we need to assume that our target function is \emph{constrained} in some way. 

Given an example behaviour like \msf{reverse} [2, 3] = [3, 2], why should we prefer the extrapolation \msf{reverse} [2, 4, 3] = [3, 4, 2] over some quus-like alternative like \msf{reverse} [2, 4, 3] = [2, 4, 3]? We identify cases where this question has a precise answer based on certain symmetries, which does not rely on any implementation or implied semantics of \msf{reverse}. In general, we will define and study a class of list functions whose action on longer lists is precisely determined by their action on shorter ones. 

% Specifically, here we will define a class of list functions whose action on longer lists is precisely determined by their action on shorter ones. 

Our contribution is mathematical, rather than philosophical, and we certainly do not claim to have solved the age-old philosophical problem of induction (our symmetries themselves could be subject to skeptical regress, for example), but we do present a novel geometric approach to length-based generalisation that makes precise and quantitative a special case of the generally-puzzling question of rule-based extrapolation. 

A second, practical, motivation for our work comes from artificial intelligence and machine learning. In 2025, it is difficult to find tasks that neural network models do not do well, but length-based generalisation remains one such case. For example, one can train a neural network model to operate on lists of length up to $20$, but still expect to see failures of generalisation at length $20,000$, or even before. In many other domains – image, audio, etc – symmetries have been a crucial tool in supporting out-of-distribution generalisation. We do not develop this line of work here, but in future, the symmetries we identify for list functions could play a similar role for length generalisation of neural networks. 

\section{Symmetries of list functions}

Now we are ready to study symmetries of list functions, i.e. functions of the form $f : [a]\rightarrow b$, mapping lists of elements of type $a$ -- which we denote using $[a]$ -- to an output of type $b$.

As will be shown, the specific kind of symmetric behaviour we will study here -- which we call \emph{equivariance} -- necessarily implies that $b = [a]$, i.e., the function must map from lists to lists, without changing the underlying type of the elements. This allows the composition of these functions in arbitrary order. We define list function equivariance as follows:
\begin{definition}
We say that a list function $f : [a]\rightarrow[a]$ is \textbf{equivariant} with respect to a particular transformation $g : [a]\rightarrow[a]$ if, when composing $f$ with $g$, the order of composition does not matter:
\begin{equation}
    g \cdot f = f \cdot g
\end{equation}
\end{definition}
Readers familiar with geometric deep learning \cite[GDL]{bronstein2021geometric} will likely recognise that this definition of equivariance is \emph{weaker} than usual, as $g$ is not constrained by a \emph{group} structure. This means it does not need to be a \emph{lossless} transformation. In this sense, the functions we study here are more easily expressed by frameworks going beyond GDL, such as categorical deep learning \cite[CDL]{pmlr-v235-gavranovic24a}.

\subsection{Equivariance to \msf{map}}

Our work is not the first to study list function symmetries. In particular, a well studied symmetry is with respect to the $\msf{map} : (a\rightarrow b) \rightarrow [a]\rightarrow[b]$ function. For a given element-wise function $\psi : a\rightarrow b$, $\msf{map}\ \psi : [a]\rightarrow[b]$ applies $\psi$ to each element of the input independently. For example:
\begin{align*}
    \msf{map}\ (+1)\ [1, 2, 3] &= [2, 3, 4]\\
    \msf{map}\ (\times 2)\ [1, 2, 3, 4] &= [2, 4, 6, 8]
\end{align*}
Since the exact list function induced by \msf{map} depends on the choice of the function $\psi$, a function that is truly symmetric to \msf{map} must remain equivariant across \emph{all such functions}:
\begin{definition}
We say that a list function $f : [a]\rightarrow[a]$ is \msf{map}-\textbf{equivariant} if, for all choices of $\psi : a\rightarrow b$, composing $f$ with $\msf{map}\ \psi$ gives the same result in either order:
\begin{equation}
    \forall\psi : a\rightarrow b.\ (\msf{map}\ \psi) \cdot f = f \cdot (\msf{map}\ \psi)
\end{equation}
\end{definition}
Such functions operate independently of the specific values of their input elements. One standard example is the \emph{reversal} function, e.g.:
\begin{equation}
    \msf{reverse}\ [1, 2, 3] = [3, 2, 1]
\end{equation}
It is simple to check that \msf{reverse} is \msf{map}-equivariant. Since it only accesses and modifies the \emph{positions} of list elements -- not their values -- any element-wise transformations of these values leave the action of \msf{reverse} unchanged.

Mathematically, it is a standard result that we can characterise all functions that commute with \msf{map} as \emph{natural transformations} from the list functor to itself, and the idea may be generalized to other functors too, such as trees. From a type theory point of view, these functions are \emph{parametrically polymorphic}, having type $[a] \to [a]$, where $a$ is a universally quantified type variable.

Returning briefly to our machine learning motivation, \msf{map} symmetries have been used to improve the data efficiency of neural sequence models, exploiting the fact that any single input-output example is informationally equivalent to the whole class of examples obtained from it by applying \msf{map} \cite{lewis2019data, Gordon2020Permutation}.

Thus, \msf{map} symmetries encode a semantically interesting property of list functions (element-identity invariance), and yield an elegant mathematical formalism (natural transformations), and a precise encoding in type theory (parametric polymorphism\!\!~\footnote{We note that parametric polymorphism is a significantly more general theory covering, for example, mixed variance type constructors.}).

While $\msf{map}$-equivariant functions are certainly important and elegant, they have no direct bearing on length-general extrapolation; as $\msf{map}$ does not change the length of the list it operates over, the $\msf{map}$ equivariance condition only specifies predictable behaviours across lists of identical lengths---not across different length ones. Exploring such symmetries will be the focus of the remainder of our paper.

\subsection{How to specify length-general symmetries?}

For a list function $f$, \msf{map} equivariance asserts that $f$ does not depend on its input elements' values by encoding the fact that the effect of $f$ is unchanged by changing these values---and values are changed precisely by applying $\msf{map}$. If we want to apply similar principles to design an equivariance condition for length-general behaviour, we need a family of \emph{length perturbation} operators. 

We can perturb a list's length either by adding or removing elements from it. A few examples of candidate perturbations show some of the degrees of freedom in specifying such a perturbation, and illustrate the fact that many choices of perturbation family do \emph{not} describe that length generalization properties we are interested in. 

% First we will describe two simple strategies to add or remove elements, to build our intuition but also to motivate the choice we make of removing elements by value.  

Specifying how to add elements can be challenging, because it is not possible to add a generic element at a generic position; some specific element must be chosen and added at some specific position. 
As an example, we will briefly look at  adding a particular element in the last position. Consider $f : [a] \to [a]$ for some fixed type variable $a$, choose some fixed $c : a$, and consider the symmetry:
\begin{equation}\label{elt_symmetry}
\forall xs : [a].\ f (xs \mdoubleplus [c]) = (f\ xs) \mdoubleplus [c],
\end{equation}
where $\mdoubleplus : [a] \times [a]\rightarrow [a]$ is list concatenation.
This equivariance constraint specifies that adding $c$ to the end of the list may be done either before or after applying $f$ to it, with unchanged results. However, it only becomes active once a list ends in $c$; on any other list, the function can behave arbitrarily. Thus, this constraint is too weak for our purposes, and it does not meaningfully constrain a function class in a way that could be relevant to extrapolation. 

Now, consider what happens if instead we focus on \emph{reducing} a list's length by removing an element. For example, consider the \msf{tail} function, which removes the first element of a list; e.g.:
\begin{equation*}
    \msf{tail}\ [1, 2, 3] = [2, 3]
\end{equation*}
Equivariance with respect to this function amounts to the following condition:
\[
f \cdot \msf{tail} = \msf{tail} \cdot f
\]
It is simple to verify that the class of \msf{tail}-equivariant functions includes \msf{map}, functions that append an arbitrary list to the end of their input, and functions that drop some number of elements from the front of their input. But it is also easy to verify that it does \emph{not} include \msf{reverse}, \msf{sort}, and many other commonly-used functions, which clearly behave predictably at smaller lengths. We can conclude that \msf{tail} equivariance is not expressive enough to characterise many extrapolating functions of interest.

A key reason why \msf{tail} was not sufficiently expressive is because it removes elements by \emph{position}---specifically, it always removes the first element. Any function doing so would not commute with \emph{permuting} functions like \msf{reverse} or \msf{sort}, as they will modify which element lands in which position, and hence which element gets removed, hindering the equivariance condition. 

However, equivariances with respect to functions removing elements by \emph{value} include many functions of interest like \msf{reverse} and \msf{sort}. We hypothesise that this equivariance is rich and interesting enough, and focus on it for the rest of the paper.
%Accordingly, we conclude that we must focus on equivariance with respect to functions removing elements by \emph{value}. 
As we will see, this will lead to a symmetry class of length-invariant functions and can offer one starting point of understanding and formalizing length generalisation. 
%\emph{does} define an interesting symmetry class of length-invariant functions, including the aforementioned \msf{reverse} and \msf{sort}, as well as several other common functions. 
\begin{figure}
    \centering
    \includegraphics[width=0.8\linewidth]{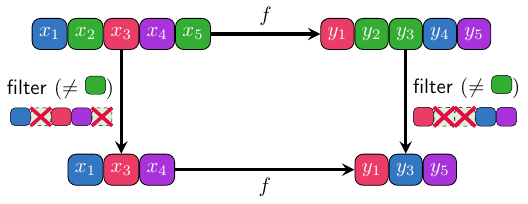}
    \caption{We leverage \msf{filter} equivariance as a way to express functions which are length-invariant through removing items by value. Generally, a \msf{filter}-equivariant (FE) function $f$ may be composed with \msf{filter} in any order, yielding the same results at the end.}
    \label{fig:fe}
\end{figure}

\subsection{Equivariance to \msf{filter}}

The canonical way to remove elements by value is to apply the function $\msf{filter} : (a \to \msf{Bool}) \to [a] \to [a]$ that selects just those elements from a list that match a predicate function $\phi : a\rightarrow\msf{Bool}$; for example:
\begin{align*}
\msf{filter}\ \msf{even}\ [1,2,3] &= [2]\\
\msf{filter}\ \msf{odd}\ [1, 2, 3, 4] &= [1, 3]
\end{align*}
Just as in the case of \msf{map}, for a function to be symmetric to all kinds of value-based removals, we need it to remain equivariant across all choices of predicates $\phi$, as follows (see Figure \ref{fig:fe}):
\begin{definition}
We say that a list function $f : [a]\rightarrow[a]$ is \msf{filter}-\textbf{equivariant} (FE) if, for all choices of $\phi : a\rightarrow \msf{Bool}$, composing $f$ with $\msf{filter}\ \phi$ gives the same result in either order:
\begin{equation}
    \forall\phi : a\rightarrow \msf{Bool}.\ (\msf{filter}\ \phi) \cdot f = f \cdot (\msf{filter}\ \phi)
\end{equation}
Further, $f : [a]\rightarrow [a]$ is \textbf{natural} \msf{filter}-\textbf{equivariant} (NFE) if it is both \msf{map-} and \msf{filter-}equivariant.
\end{definition}
We will make precise the sense in which \msf{filter} equivariance supports extrapolation by showing that, with one additional technical requirement, \msf{filter}-equivariant functions are fully determined by their behavior on lists with \textbf{two unique elements}. Further, natural \msf{filter} equivariants are determined by a \textbf{single} length-two example. For example, knowing that \msf{reverse} [2, 3] = [3, 2], and that \msf{reverse} is both \msf{map}- and \msf{filter}-equivariant, suffices to determine the value of \msf{reverse} for \emph{any} other input list!

In what follows, we will study specific examples of (N)FEs and leverage them to build a better overall intuition for how they operate. This will culminate with the \emph{amalgamation} algorithm in Section 5, which will allow us to construct all FEs from knowledge of their behaviour on length-two sublists.

%We define an inductive datatype for natural filter invariants, enumerate all such functions explicitly, and present a model of filter invariants in semi-simplicial geometry in which filter invariants emerge as natural transformations. We also prove a theorem concerning when recursive functions are filter-invariant. 

% One might even use this to postulate that rule-following is naturality, with the only question being naturality between what functors.

Before we proceed, a brief note on our usage of the word `symmetry'. Usually, symmetries are defined in terms of the action of a group of transformations. However, neither of the operations discussed in this paper (\msf{map} and \msf{filter}) form groups, as they are not invertible. Thus, the symmetries in this paper are with respect to a \emph{monoid} rather than a group. This will not prove to be a limitation for deriving our results, and hence we will not reference this point further. That said, it presents challenges for building these symmetries into neural networks using established techniques from geometric deep learning. More advanced techniques, exemplified by works like \cite{pmlr-v231-dudzik24a}, are likely to be required.

\section{Analysing (natural) \msf{filter} equivariants}

Here are some examples of (N)FEs worth considering:
\begin{itemize}
    \item The \msf{reverse} function is an NFE.
    \item Another NFE is the $\msf{inflate} \ n$ function, where
    $\msf{inflate} : \msf{Nat} \to [a] \to [a]$ is defined by 
    \[
\msf{inflate} \ n = \msf{concat} \cdot \msf{map} \ (\msf{repeat} \ n),
\]
and $\msf{repeat} : \msf{Nat}\rightarrow a\rightarrow [a]$ is implemented as:
\begin{equation*}
\msf{repeat}\ n \ x = [\ \underbrace{x, x, \dots, x}_{n\ \text{times}}\ ].
\end{equation*} 
Thus $\msf{repeat}\ n\ x$ is a list containing $n$ copies of $x$, and $\msf{inflate}\ n$ repeats each element in the list $n$ times, for example:
\begin{equation*}
\msf{inflate}\ 3\ [1, 2, 3] = [1, 1, 1, 2, 2, 2, 3, 3, 3].
\end{equation*}

    \item FEs that are not (always) NFE include the \msf{sort} function, as well as $\msf{filter}\ \phi$ itself. 
    \item An example of a function that is \msf{map}-equivariant, but is \emph{not} NFE is \msf{triangle}, which repeats the $i$th element of an input list $i$ times, e.g.:
    \[
    \msf{triangle } [3, 7, 5] = [3, 7, 7, 5, 5, 5]
    \]
\end{itemize}

Studying the above examples, we can develop a mental model of what \msf{filter}-equivariant functions look like. Just as \msf{map} equivariants cannot depend on elements' values, \msf{filter} equivariants cannot depend on elements' \emph{absolute position}. This is an intuitive constraint, because the application of \msf{filter} may easily displace absolute positions. Functions that depend on the elements' \emph{relative} positions -- like \msf{reverse} -- or that are completely independent of the input order -- like \msf{sort}\footnote{Note that \msf{sort} is only completely \emph{independent} of the input order if its comparison function is not antisymmetric.} -- both meet this criterion. In the non-example \msf{triangle}, an element's output multiplicity depends on its absolute input position.

Beyond these examples, we can create new (N)FEs by combining existing ones via \emph{concatenation} and \emph{composition}. In other words, (N)FEs possess the following two monoid structures:
\begin{lemma}
    NFEs (and FEs) form a monoid with unit being the constant function mapping every list to the empty list, and with addition defined by pointwise list concatenation:
    \[
(f \mdoubleplus g) \; xs = (f \; xs) \mdoubleplus (g \; xs)
    \]
\end{lemma}
\begin{lemma}
    NFEs (and FEs) form another monoid with the unit being the identity function and the composition of two (N)FEs giving an (N)FE.
\end{lemma}
Given a list $xs :: [a]$, we denote by $|xs|$ the elements occurring in $xs$. This function may be defined, for example, in Haskell, as follows:

\begin{Verbatim}[commandchars=\\\{\}]
\PYG{+w}{    }\PYG{o}{|\PYGZhy{}|}\PYG{+w}{ }\PYG{o+ow}{::}\PYG{+w}{ }\PYG{p}{[}\PYG{n}{a}\PYG{p}{]}\PYG{+w}{ }\PYG{o+ow}{\PYGZhy{}\PYGZgt{}}\PYG{+w}{ }\PYG{p}{[}\PYG{n}{a}\PYG{p}{]}
\PYG{+w}{    }\PYG{o}{|}\PYG{+w}{ }\PYG{k+kt}{[]}\PYG{+w}{ }\PYG{o}{|}\PYG{+w}{ }\PYG{o+ow}{=}\PYG{+w}{ }\PYG{k+kt}{[]}
\PYG{+w}{    }\PYG{o}{|}\PYG{n}{x}\PYG{k+kt}{:}\PYG{n}{xs}\PYG{o}{|}\PYG{+w}{ }\PYG{o+ow}{=}\PYG{+w}{ }\PYG{n}{x}\PYG{+w}{ }\PYG{k+kt}{:}\PYG{+w}{ }\PYG{o}{|}\PYG{+w}{ }\PYG{n}{filter}\PYG{+w}{ }\PYG{p}{(}\PYG{o}{/=}\PYG{+w}{ }\PYG{n}{x}\PYG{p}{)}\PYG{+w}{ }\PYG{n}{xs}\PYG{+w}{ }\PYG{o}{|}
\end{Verbatim}

Note that $|-| : [a]\rightarrow [a]$ is FE but not NFE. Our first insight is that FE functions do not add any new values to the input list. (As a corollary, FE functions must map the empty list to the empty list.)
\begin{lemma}\label{lem:nonew}
Let $f : [a] \to [a]$ be \msf{filter}-equivariant. Then, for all input lists 
$xs : [a]$, $|f \; xs| \subseteq |xs|$.
\end{lemma}
{\bf Proof:} Let $xs$ be a list, assume  $y\notin |xs|$ and we can show $y\notin |f \; xs|$. Define the predicate $\phi$ by $\phi\ z := (z \neq y)$. Then $f\ xs = (f\cdot \msf{filter}\ \phi) \ xs = (\msf{filter}\ \phi\cdot f)\ xs = \msf{filter}\ \phi\ (f\ xs)$. Therefore, $y\notin |f \ xs|$.

\subsection{Properties of ($k$-)NFE functions}\label{subsec:nfes}
Recall that natural \msf{filter} equivariants (NFEs) are functions that are equivariant to both \msf{map} and \msf{filter}. This is a very strong pair of constraints: it implies that the function must both behave predictably across various lengths \emph{and} its behaviour must not depend on the values of the items inside the list. We begin by studying NFEs, as these constraints will more easily yield interesting mathematical structure.

As a general note for what follows, we may assume without loss of generality that the input list to any NFE has unique elements. One way to force the elements of an input list $xs$ to be unique by ``tagging'' each element with its index: 
\[
\msf{enumerate}\ [x_0, x_1, \ldots, x_n] = [(x_0, 0), (x_1, 1), \ldots, (x_n, n)]
\]
This operation can be undone by:
\[
\msf{unenumerate}\ [(x_0, 0), (x_1, 1), \ldots, (x_n, n)] = [x_0, x_1, \ldots, x_n]
\]

Noting that $\msf{unenumerate} = \msf{map } \msf{first}$, we can see that uniquification of input elements preserves the operation of any natural transformation $f$, in the following sense:
\[
\begin{aligned}
\msf{unenumerate} \cdot f \cdot \msf{enumerate} 
  &= (\msf{map}\;\msf{first}) \cdot f \cdot \msf{enumerate} \\[4pt]
  &= f \cdot (\msf{map}\;\msf{first}) \cdot \msf{enumerate} \\[4pt]
  &= f \cdot \msf{unenumerate} \cdot \msf{enumerate} \\[4pt]
  &= f
\end{aligned}
\]

% Such lists may then be further processed. Any list processed by \msf{enumerate} may then be rolled back:
% \[
% \msf{unenumerate}\ [(x_0, 0), (x_1, 1), \ldots, (x_n, n)] = [x_0, x_1, \ldots, x_n]
% \]
% This enables us to define functions that behave well with respect to uniquification:
% \[
% f \cdot \msf{enumerate} = \msf{unenumerate} \cdot f
% \]

The first result we show for NFEs is that they must replicate each element a consistent number of times---if a particular element is replicated $k$ times by an NFE, \emph{all} of them must be replicated $k$ times.
\begin{lemma}\label{lem:counts}
    Let $f:[a] \to [a]$ be natural \msf{filter}-equivariant and let $k = \msf{len}(f \; [\ast])$ where $\ast:1$. Then, for all inputs $xs : [a]$,  we have $\msf{len} \; (f \; xs) = k \times (\msf{len} \; xs)$. Further, if $x$ occurs $m$ times in $xs$, then $x$ occurs $k \times m$ times in $f \; xs$.
\end{lemma}
{\bf Proof:} First, we use the \msf{map} equivariance of $f$ to show that, when applied to any singleton list, it must return a list of length $k$. Further, we know -- due to Lemma \ref{lem:nonew} -- that this list can only contain the singleton element in the original input list. Now, we can use the \msf{filter} equivariance of $f$ to extend the result for singletons to arbitrary lists with no repeated elements. For the case where the input list contains repeated elements, we can apply \msf{enumerate} as described above, apply the previous result, and then use \msf{map} equivariance to conclude the proof. 

We call a function $f: [a] \to [a]$ a $k$-NFE if it is an NFE that maps singletons to lists of length $k$. We first describe 1-NFEs and then tackle the general case. Since Lemma \ref{lem:counts} implies that 1-NFEs do not modify the overall collection of values in the input, they are nothing more than a pre-defined \emph{permutation} of the input elements---a standard example of which is \msf{reverse}. By \msf{map} equivariance, a 1-NFE must therefore be given by a family of permutations $t : \prod_{n : \msf{Nat}}\Perm(n)$ where $\Perm$ maps every natural number $n$ to the set of permutations on the set $\{0, .., n-1\}$. For the specific case of \msf{reverse}, these permutations are
\begin{equation*}
    t_{\msf{reverse}} = [[], [0], [1, 0], [2, 1, 0], [3, 2, 1, 0]\dots]
\end{equation*}
While this approach will allow us to define 1-NFEs, we still need to include the coherence conditions relating the various permutations ($t\ 0$ to $t\ 1$ to $t\ 2$, etc.). This is the essence of length-generality: the permutations at longer lengths need to be systematically related to the ones at shorter lengths. To define this coherence, we leverage simplicial algebra:
\begin{definition}
    Let $\Delta$ be the semi-simplicial category. Its objects are the finite natural numbers (thought of as sets $\{0, 1, \dots, n-1\}$ for $n : \msf{Nat}$), and its morphisms $n \to m$ are inclusions. A \textbf{semi-simplical set} is a functor $F:\Delta^\mathrm{op} \to \Set$, i.e. a family of sets $F(n)$ with functions $F(f) : F(m) \to F(n)$ for every inclusion $f : n \to m$, obeying the functoriality laws.
\end{definition}
With the concept of semi-simplicial sets handy, we can define two particularly important semi-simplicial sets: the \emph{terminal} set and the \emph{permutation} set:
\begin{example}
    The terminal semi-simplicial set, $1$, is the functor that maps every object to the one-element set. The functor $\Perm$ is the semi-simplicial set that maps $n$ to the set $\Perm(n)$ of permutations on $n$. Given any inclusion $f : n \rightarrow m$, define $\Perm(f) : \Perm(m)\rightarrow\Perm(n)$ to map a permutation on $m$ elements to the permutation on $n$ elements, by simply removing the elements in $m$ that are not in $n$.
\end{example}
Using these two semi-simplicial sets, we can formally, coherently define a 1-NFE:
\begin{lemma}
    A 1-NFE is exactly the natural transformation 
    $1 \to \Perm$. That is, a 1-NFE is a family of permutations $t_n : \Perm(n) $ such that for every inclusion $f : n \to m$, $t_n  = \Perm(f)\ t_m$. We call such families \textbf{semi-simplicial permutations}.
\end{lemma}

Indeed, those familiar with the terminology of category theory will note that a 1-NFE is simply a \textit{cone} over the functor $\Perm$ and that the set of 1-NFEs is $\lim_{n\rightarrow\infty} \Perm(n)$. Below we will refer to families like this as cones, for brevity.

Next, we turn to the general case of $k$-NFEs, where a similar story can be told. We know -- by Lemma \ref{lem:counts} -- that a $k$-NFE maps lists of length $n$ to lists of length $k \times n$, by mapping a list $xs$ to a permutation of $\msf{inflate} \; k \; xs$. Because of the \msf{map} equivariance of NFEs, this data is given through a permutation of $k \times n$, i.e., an element of $\Perm(k \times n)$. Thus, we define:
\begin{definition}
The semi-simplicial set of $k$-permutations is defined by the functor 
\[
\Delta^\mathrm{op} \stackrel{-\times k}{\longrightarrow} \Delta^\mathrm{op}
\stackrel{\Perm}{\longrightarrow} \Set
\]
A cone for the above functor is called a \textbf{$k$-semi-simplicial permutation}.
\end{definition}
Using $k$-semi-simplicial permutations, we can extend our definition of 1-NFEs to $k$-NFEs analogously:
\begin{lemma}
    A particular $k$-NFE is a $k$-semi-simplicial permutation, and hence the set of all $k$-NFEs is $\lim_{n\to\infty} \Perm(n\times k)$.
\end{lemma}

Just as natural transformations give a categorical interpretation of \msf{map} equivariance, semi-semplicial permutations develop the category theory for \msf{filter} equivariance. In the next section, we turn to type theory. 

%This gives me the following conjecture
%
%\begin{lemma}[Conjecture]
%    Every $k$-NFI is $rpt \; k$ followed by a 1-NFI
%\end{lemma}

\subsection{Inductive characterisation of NFEs}
Given the relationship between $k$-semi-simplicial permutations and $1$-semi-simplicial permutations, one may wonder if $k$-NFEs can be built from 1-NFEs. A positive answer would hint at a compositional semantics for NFEs where $k$-NFEs are built from $k'$-NFEs for $k' \leq k$.

This claim turns out to be true, and it is possible to provide an \emph{inductive} presentation of the set of NFEs as an \emph{inductive data type}. Concretely, in Haskell, we have:

\begin{Verbatim}[commandchars=\\\{\}]
\PYG{+w}{    }\PYG{k+kr}{data}\PYG{+w}{ }\PYG{k+kt}{NFE}\PYG{+w}{ }\PYG{o+ow}{=}\PYG{+w}{ }\PYG{k+kt}{Z}
\PYG{+w}{             }\PYG{o}{|}\PYG{+w}{ }\PYG{k+kt}{P}\PYG{+w}{ }\PYG{k+kt}{Nat}\PYG{+w}{ }\PYG{k+kt}{NFE}
\PYG{+w}{             }\PYG{o}{|}\PYG{+w}{ }\PYG{k+kt}{N}\PYG{+w}{ }\PYG{k+kt}{Nat}\PYG{+w}{ }\PYG{k+kt}{NFE}
\end{Verbatim}

To be clear, the data type {\tt NFE} contains only the unique representations of NFEs~\footnote{The constructors {\tt P} and {\tt N} are assumed to have their first input as $n : \msf{Nat}$, where $n \geq 1$.} and is a drastic simplification of the \emph{formal} definition of NFEs, which consists of data (as above) which must obey equivariance constraints (missing from the above). 

Rather than providing NFEs as an inductive data type, we could give $k$-NFEs a definition as an inductive family, by defining $\msf{NFE} : \Nat \to \Set$ mapping $k$ to the set of $k$-NFEs. The only reason why we chose not to do so is because we have currently found no use for such a presentation. 

To understand the inductive type {\tt NFE}, consider it as a defining lists of natural numbers with two different ways to construct them. Indeed, 
\begin{lemma}
    There is a bijection between $\msf{NFE}$ and $\msf{List}(\Nat + \Nat)$.
\end{lemma}
We now explain how {\tt NFE} represents an NFE by giving a function mapping every element of the data type {\tt NFE} to an actual NFE. Since {\tt NFE} is the free monad on $\msf{Nat} + \msf{Nat}$, and NFEs form a monoid, it suffices to map $\msf{Nat} + \msf{Nat}$ to the class of NFEs. We do this by sending an element $n$ of the first injection to the NFE $\msf{inflate}\ n$ and an element $n$ of the second injection to $\msf{reverse} \cdot (\msf{inflate}\ n)$. This gives the following function

\begin{Verbatim}[commandchars=\\\{\}]
\PYG{+w}{    }\PYG{p}{[[}\PYG{+w}{ }\PYG{o}{\PYGZhy{}}\PYG{+w}{ }\PYG{p}{]]}\PYG{+w}{ }\PYG{o+ow}{::}\PYG{+w}{ }\PYG{k+kt}{NFE}\PYG{+w}{ }\PYG{o+ow}{\PYGZhy{}\PYGZgt{}}\PYG{+w}{ }\PYG{p}{[}\PYG{n}{a}\PYG{p}{]}\PYG{+w}{ }\PYG{o+ow}{\PYGZhy{}\PYGZgt{}}\PYG{+w}{ }\PYG{p}{[}\PYG{n}{a}\PYG{p}{]}
\PYG{+w}{    }\PYG{p}{[[}\PYG{+w}{ }\PYG{k+kt}{Z}\PYG{+w}{ }\PYG{p}{]]}\PYG{+w}{ }\PYG{n}{xs}\PYG{+w}{     }\PYG{o+ow}{=}\PYG{+w}{ }\PYG{k+kt}{[]}
\PYG{+w}{    }\PYG{p}{[[}\PYG{+w}{ }\PYG{k+kt}{P}\PYG{+w}{ }\PYG{n}{n}\PYG{+w}{ }\PYG{n}{m}\PYG{+w}{ }\PYG{p}{]]}\PYG{+w}{ }\PYG{n}{xs}\PYG{+w}{ }\PYG{o+ow}{=}\PYG{+w}{ }\PYG{n}{inflate}\PYG{+w}{ }\PYG{n}{n}\PYG{+w}{ }\PYG{n}{xs}\PYG{+w}{ }\PYG{o}{++}\PYG{+w}{ }\PYG{p}{[[}\PYG{+w}{ }\PYG{n}{m}\PYG{+w}{ }\PYG{p}{]]}\PYG{+w}{ }\PYG{n}{xs}
\PYG{+w}{    }\PYG{p}{[[}\PYG{+w}{ }\PYG{k+kt}{N}\PYG{+w}{ }\PYG{n}{n}\PYG{+w}{ }\PYG{n}{m}\PYG{+w}{ }\PYG{p}{]]}\PYG{+w}{ }\PYG{n}{xs}\PYG{+w}{ }\PYG{o+ow}{=}\PYG{+w}{ }\PYG{n}{reverse}\PYG{+w}{ }\PYG{p}{(}\PYG{n}{inflate}\PYG{+w}{ }\PYG{n}{n}\PYG{+w}{ }\PYG{n}{xs}\PYG{p}{)}\PYG{+w}{ }\PYG{o}{++}\PYG{+w}{ }\PYG{p}{[[}\PYG{+w}{ }\PYG{n}{m}\PYG{+w}{ }\PYG{p}{]]}\PYG{+w}{ }\PYG{n}{xs}
\end{Verbatim}

That every element of {\tt NFE} generates an NFE is clear, since we have already seen that NFEs contain the constant function returning the empty list {\tt []}, contain $\msf{reverse}$ and $\msf{inflate} \ n$, and are closed under function composition and under concatenation. To see these are the \emph{only} NFEs, we can first leverage \msf{map} equivariance to show that the only 1-NFEs are the \msf{identity} and \msf{reverse}. 

This result scales to $k$-NFEs by enumeration across $k$; it further shows the following:
\begin{lemma}
    The number of $k$-NFEs is $2\times 3^{k-1}$ when $k \geq 1$. 
\end{lemma}
For example, the only 2-NFEs are $\msf{inflate}\ 2$, $\msf{reverse} \cdot (\msf{inflate} \ 2)$, $\msf{identity} \mdoubleplus \msf{identity}$, $\msf{reverse} \mdoubleplus \msf{identity}$, $\msf{identity} \mdoubleplus\msf{reverse}$, and $\msf{reverse}\mdoubleplus\msf{reverse}$ -- six functions in total.

\section{Characterising \msf{filter} equivariants}

Having established the foundations that allowed us to characterise \emph{all} NFEs, we are ready to relax the \msf{map} equivariance constraint, and refine our analysis for the case of FEs -- the crux of our paper.

Recall how we characterised NFEs by the fact they will \emph{inflate} the input list by a certain constant $k$ (leading to $k$-NFEs). It is possible to generalise this to the FE case---however, when doing so, not all elements will necessarily be replicated by the same amount, leading to the following result:
\begin{lemma}
    Given a \msf{filter}-equivariant function $f:[a] \to [a]$, there is an underlying function $\Phi: a \to \Nat \to \Nat$ such that if $x$ occurs $n$ times in $xs$, then $x$ occurs $\Phi(x,n)$ times in $f\ xs$
\end{lemma}
{\bf Proof:} The proof of the above defines
\[
\Phi(x,n) = \msf{len} (f\ (\msf{repeat} \ n \ x))
\]
To prove the lemma, use \msf{filter} equivariance for the specific function $\msf{filter}\ \phi$ keeping only the element of interest, i.e., $\phi\ z := (z=x)$. \hfill $\Box$

With knowledge of the occurrence function $\Phi$, we can characterise the final output of any FE as follows:
\begin{lemma}
    Let $f$ be \msf{filter}-equivariant. Then, if the distinct elements of $xs$ are
    $[x_1, \ldots, x_n]$ and $x_i$ occurs $n_i$ times in $xs$, then $f\ xs$ is a permutation of 
    \[
\msf{concat} \; [ \; \msf{repeat} \; \Phi(x_i, n_i) \; x_i \;\; | \;\; 1 \leq i \leq n\ ]
    \]
\end{lemma}
These permutations must be coherent with respect to the same semi-simplicial structure used to describe NFEs. To formalise this statement, we can generalise the idea of $k$-NFEs to $\Phi$-FEs, where $\Phi : a\to\Nat\to\Nat$ is the corresponding occurrence function of the FE.

We characterised $k$-NFEs via functors whose domain was the semi-simplicial category $\Delta$, because all we needed to know was the length of the list -- \msf{map} equivariance meant we could assume, without loss of generality, that all elements were distinct. This is not the case for $\Phi$-FEs where we need to know \emph{which elements are in the input list} and \emph{what is their multiplicity} (the two inputs for $\Phi$). This is a setting suitable for \emph{multisets} (also known as \emph{bags}), and hence we define category $\msf{Bag}$ as follows
\begin{definition}
    The category $\msf{Bag}$ is defined as follows
    \begin{itemize}
        \item Objects are tuples $(n, f)$ with finite numbers $n : \Nat$ and functions $f:\Fin\; n \to \Nat$.\footnote{Here $\msf{Fin} : \msf{Nat} \to \Set$ explicitly transforms a number to a set of that size.}
        \item Morphisms $(n,f)\to(n',f')$ are inclusions $i:n \to n'$ such that $f = f' \cdot i$  
    \end{itemize}
\end{definition}
The idea is that an object $(n,f)$ represents a multiset of $n$ distinct elements, where the $i$th element ($0\leq i\leq n - 1$) occurs $f\ i$ times. Using this structure -- which entirely represents our input list's item counts -- we can now map it into the semi-simplicial category $\Delta$:
\begin{definition}
    Given an occurrence function $\Phi$ as above, there is an induced functor $\hat{\Phi} : \msf{Bag} \to \Delta$ sending $(n,f)$ to 
    $(\sum_i : \Fin \ n) \; \Phi(i, f\ i)$.
\end{definition}
Now we can use this mapping to characterise all $\Phi$-FEs, just as we did for $k$-NFEs before. A $\Phi$-FE will be a family of permutations which are coherent to ensure \msf{filter} equivariance. This is exactly a $\hat{\Phi}$-cone.
\begin{lemma}
    A $\Phi$-FE function is a cone over the functor $\Perm \cdot \hat{\Phi}$. 
\end{lemma}
What is pleasing about this construction is that we have a very similar characterisation for $\Phi$-FEs as we had for $k$-NFEs. Indeed, the characterisations  coincide on NFEs. Using this, we can
enumerate specific mechanisms for defining \msf{filter}-equivariant functions. The most interesting is the clause showing how \emph{recursive} functions defined by \emph{iteration} may be \msf{filter} equivariant.
\begin{lemma} The following claims are all true:
\begin{itemize}
    \item If $f \in \msf{FE}$ and $f' \in \msf{FE}$, then $f \mdoubleplus f' \in \msf{FE}$ and $f\cdot f' \in \msf{FE}$.
    \item If $f \in \msf{NFE}$ then $f \in \msf{FE}$.
    \item If $\alpha : a \to [a] \to [a]$ is such that 
\begin{eqnarray*} 
\msf{filter} \; \phi \; (\alpha (x, xs)) &  =  & 
\msf{filter} \; \phi \; xs \;\;\;\;\;\;\;\;\;\;\;\;\;\; \phi\ x = \msf{false}\\
\msf{filter} \; \phi \; (\alpha (x, xs)) &  =  & 
\alpha (x, \msf{filter} \; \phi \; xs) \;\;\;\;\; \phi\ x = \msf{true}\\
\end{eqnarray*}
then $\msf{foldr} \; \alpha \; [] \; \in \; \msf{FE}$, where \msf{foldr} is the \emph{right fold}\footnote{The general signature of the right fold is $\msf{foldr} : (a \to b \to b) \to b \to [a] \to b$. $\msf{foldr}\ f\ z : [a]\rightarrow b$ is a standard way to represent iterative computation: the function $f$ is applied iteratively to the entries of the input list in $[a]$, with the value of $z : b$ used to ``seed'' the computation.}, such that, for a given list $xs = [x_1,\dots,x_n]$:
\begin{equation*}\msf{foldr}\ \alpha\ []\ xs = \alpha \left(x_1, \alpha\left(x_2, \alpha(\dots,\alpha(x_{n-1},\alpha(x_n, [])))\right)\right).
\end{equation*}
Note this example covers \msf{sort}, \msf{reverse} and $\msf{filter} \; \phi$, most of the specific FEs we discussed before.
\end{itemize}
\end{lemma}

\section{Amalgamation}

In this section, we ``cash out'' the previous definitions and results, and show exactly how filter equivariance supports length generalisation. In particular, we show how an NFE's value on \emph{any} input is determined by its value on a single length-two input, and an FE's behavior on any list is determined by all of that list's sublists with two unique elements.

Translating these statements into examples, if it is given that
\begin{itemize}
    \item A function $f$ is NFE,
    \item $f\ [1, 2] = [2, 1, 2, 1]$,
\end{itemize}
then we can exactly deduce $f$'s value on any other input. (In this specific case, $f = \msf{reverse} \mdoubleplus \msf{reverse}$.)

Similarly, if it is given that
\begin{itemize}
    \item A function $f$ is FE,
    \item $f\ [2, 1, 2] = [1, 2, 2]$,
    \item $f\ [3, 2, 2] = [2, 2, 3]$,
    \item $f\ [3, 1] = [1, 3]$,
\end{itemize}
then we can deduce $f\ [3, 2, 1, 2] = [1, 2, 2, 3]$. (Note that $[2, 1, 2]$, $[3, 2, 2]$ and $[3, 1]$ are all the sublists of $[3, 2, 1, 2]$ with two unique elements kept.) 

First, we'll prove this claim for the FE case, and then show how the NFE case follows as a corollary. 

\begin{figure}
    \includegraphics[width=\linewidth]{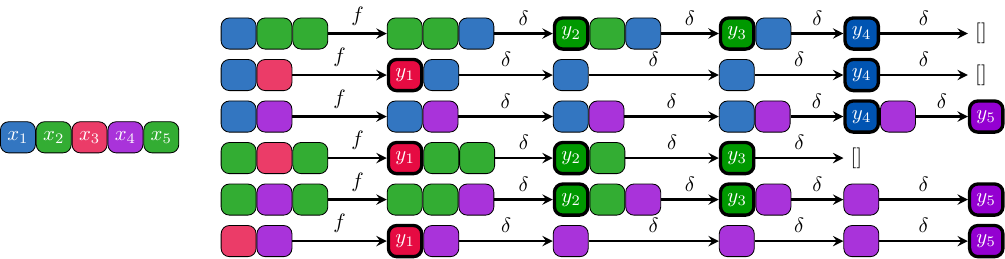}
    \caption{The process of amalgamating the output of $f\ [x_1, x_2, x_3, x_4, x_5]$, for the same \msf{filter}-equivariant function $f$ and input as presented in Figure \ref{fig:fe}. We show that it is sufficient to invoke $f$ on all sublists of two unique values (in this case, there are six such sublists), and then execute the amalgamation algorithm, \msf{amal}, over the resulting collection. At each step, the algorithm proceeds by ``majority voting'' on the first position across the collection to decide the next output in the sequence $[y_1, y_2, y_3, y_4, y_5]$. Then, using the $\delta$ function, all the picked $y_i$ elements are removed from their corresponding lists.}
    \label{fig:amal}
\end{figure}

In outline, the process to compute the value for an unseen, longer, input list $xs$ will be to apply all necessary filters to reduce $xs$ to have only two unique elements. We then compute or lookup $f$'s values on these smaller lists, and `glue' the results together. This process of gluing partial results together we call \textbf{amalgamation}, and it is depicted in Figure \ref{fig:amal}.

% This section goes back to the problem mentioned at the beginning of this paper, namely generalisation. As a concrete example, if we know the following
% \[
% \msf{reverse} \; [1,2] \;\;\;  \;\;\; \msf{reverse} \;  [1,4,4] \;\;\; \;\;\; \msf{reverse} \;[2,4,4]
% \]
% then can we automatically infer $\msf{reverse} \; [1,2,4,4]$. 

Let's start by abstracting the problem. We want to compute 
$f \; xs$ for some \msf{filter}-equivariant function $f :[a] \to [a]$ and list $xs : [a]$. To do so, we look at each element $x:a$ in $xs$, and compute
$f$ on $\msf{filter} \; (\neq x) \; xs$. A first question is: can we compute $f \; xs$ from these partial results?

To be more formal, let us introduce some additional notation. Let $X$ be a set of all values in our list, and let $X \backslash x$ denote the set $X$ with $x$ removed. We define $\pi_x : [X] \to [X\backslash x]$ to be the \msf{filter} operation removing $x$, i.e., $\pi_x = \msf{filter}\ (\neq x)$. So the question now becomes: can we compute $f \; xs$ from the collection of lists $f \; (\pi_x \; xs)$ where $x$ ranges over the unique elements of $X$?

\subsection{Amalgamability of filtered lists}

Let us go one step further and look at the collection of one-element-filtered lists. That is, we \emph{collate} the functions $\pi_x$ into one function $\pi : [X] \to X \to [X\backslash x]$ defined by 
$\pi \; xs \; x = \pi_x \; xs$.

We now focus on the type $X \to [X\backslash x]$, which we can think of as a collection of lists in $[X\backslash x]$ -- one for each element $x : X$ removed. Clearly, not every such collection of lists can be amalgamated into one unique list. Therefore, we need to carve out a subtype of $X \to [X\backslash x]$ containing only {\em amalgamable} collections---collections for which only one, unique and correct way to amalgamate exists. 

We can define this subtype recursively. Firstly, we can define the predicate $\msf{AM}_0 \; X$, of collections, $\chi : X\to [X\backslash x]$, from which the first element, $x_0 : X$, is amalgamable:
\[
\chi \in \msf{AM}_0\ X \;  \Longleftrightarrow \; 
\exists x_0 \in X.\ \forall x \neq x_0 \in X.\ \msf{head}\ (\chi \; x) = x_0,
\]
where $\msf{head} : [a]\rightarrow a$ returns the first element (``head'') of its input list.

Unpacking this definition, it means that all lists in the collection that still have $x_0$ must put it at the first position. If $X$ has at least three elements, the choice of $x_0$ must be \emph{unique}, due to the fact that there are $|X|$ sublists in $\chi$, and in $|X-1|$ of those, $x_0 = \msf{head}\ xs$ is not filtered out, and will be at the front. From now on, we therefore assume $X$ has at least three elements, and denote the head as $x_0(\chi)$.

Now, we must ensure the amalgamability is also maintained across the rest of the list, and using $\msf{AM}_0$, we can ground a recursive procedure to do so. We define a new predicate, $\msf{AM}\ X$, by:
\begin{eqnarray*}
    \chi \in \msf{AM} \; X & \Longleftrightarrow & (\forall x:X.\ \chi\ x = [])\\
    & & \vee\ (\chi \in \msf{AM}_0\ X \wedge \; \delta \chi \in \msf{AM}\; X)
\end{eqnarray*}
where $\delta \chi: X \to [X \backslash x]$ is defined as follows:
\begin{equation*}
    \delta\chi\ x =\begin{cases}
        \chi\ x & x=x_0(\chi)\\
        \msf{tail}\ (\chi\ x) & x\neq x_0(\chi)
    \end{cases}
\end{equation*}
In this definition, $x_0(\chi)$ is the unique head element of $xs$ (which must exist if $\chi\in\msf{AM}_0\ X$). In essence, the definition of $\msf{AM}\ X$ entails that amalgamable collections \emph{either} have no elements left to amalgamate, \emph{or} they can uniquely amalgamate the first element $x_0$, and the remaining collection of lists, $\delta\chi$ -- obtained by removing the head $x_0$ from all relevant lists, by the use of \msf{tail} -- is amalgamable.

\subsection{Amalgamating \msf{filter} equivariant functions from filtered sublists}

We can now build our proof. Firstly, there is an isomorphism between $[X]$ and $\msf{AM} \; X$, with one direction given by $\pi$. The inverse is the function $\msf{amal}:\msf{AM}\; X \to [X] $ given by:
\begin{equation}
    \msf{amal}\ \chi = \begin{cases}
        [] & \forall x:X.\ \chi\ x = []\\
        x_0(\chi):: \msf{amal}\ \delta\chi & \mathrm{otherwise}
    \end{cases},
\end{equation}
which also specifies an algorithm for reconstructing $xs$ given an amalgamable collection of $\pi_x\ xs$.
\begin{lemma}\label{lem:iso}
    Let $xs : [X]$ be a list with at least three unique elements. Then $\pi \; xs \in \msf{AM}\ X$. Further, $\pi$ is an isomorphism with $\msf{amal}$ as its inverse.
\end{lemma}
\noindent {\bf Proof: }  The first statement (on amalgability of $\pi\ xs$) holds by induction on $xs$, noting that $\pi \cdot \msf{tail} = \delta \cdot \pi$. This statement can also be used to prove $\msf{amal} \cdot \pi = \msf{identity}$.\hfill $\Box$

It will also be useful to note that this same logic allows us to amalgamate $\pi\ (f\ xs)$, for any \msf{filter}-equivariant $f : [X]\to [X]$:
\begin{lemma}\label{lem:f_assemble}
    Let $f : [X]\to[X]$ be a \msf{filter}-equivariant function, and $xs : |X|$ be a list with at least three unique elements. Then, $\pi\ (f\ xs)\in \msf{AM}\ X$.
\end{lemma}
{\bf Proof:}\ If $f$ does not reduce the size of $xs$ -- i.e., $|xs|=|f\ xs|$ -- then surely $f\ xs$ has at least three unique elements, and Lemma \ref{lem:iso} applies. If, instead, $|xs| > |f\ xs|$ -- for example, if $f$ is a \msf{filter} -- we need to apply more care if $|f\ xs| < 3$.

When $|f\ xs| = 0$, $f\ xs = []$, therefore $(\pi\ (f\ xs))\ x = []$ for all $x : X$, and hence $\pi\ (f\ xs)\in\msf{AM}\ X$.

Otherwise, let $\chi\ x = \pi\ (f\ xs)$ be our collection of outputs of $f$. We first show that $x_0(\chi)=\msf{head}\ (f\ xs)$. Now, since $\chi\ x = \msf{filter}\ (\ne x)\ (f\ xs)$, we can conclude that $\msf{head}\ (\chi\ x) = \msf{head}\ (f\ xs)$ whenever $x\ne\msf{head}\ (f\ xs)$. And since $|X|$ -- the number of collections in $\chi$ -- is at least 3, this element is uniquely specified. Any other candidate $x'$ for $x_0(\chi)$ would not be the head of the list $\chi\ z$, for $z\ne x',z\ne x$ (which must exist since $|X|\geq 3$). Therefore, $\pi\ (f\ xs)\in\msf{AM}_0\ X$

It is possible to show (just like for Lemma \ref{lem:iso}) that $\pi\cdot\msf{tail}=\delta\cdot\pi$, and we are done. \hfill $\Box$

Now we can use this to show our key result:
\begin{theorem} 
Let $f:[X] \to [X]$ be \msf{filter}-equivariant, and $xs : [X]$ be a list with at least three unique elements. Then, $\msf{amal}$ can perfectly reconstruct $f\ xs$ from knowledge of the collection $\{f\ ys\}$, where $ys$ ranges over all sublists of $xs$ with two unique elements. 
\end{theorem}
{\bf Proof:} First, we show that we can correctly amalgamate $f\ xs$ from the collection of all outputs of $f$ where one input element had been filtered out, i.e., $\{f\ (\msf{filter}\ (\neq x)\ xs)\}_{x : X}$:
\[
f \; xs = \msf{amal}\ (\pi\ (f \; xs)) = \msf{amal}\  (\lambda x
 \to \msf{filter} \; (\neq x) \; (f \; xs)) = \msf{amal}\ (\lambda x
 \to f\ (\msf{filter} \; (\neq x) \; xs))
 \]
where we exploited Lemma \ref{lem:f_assemble} in the first step, and the fact that $f$ is \msf{filter}-equivariant in the final step.

Now we observe that, if $\msf{filter}\ (\ne x)\ xs$ has more than two unique elements, those can be themselves reconstructed from collections obtained by filtering an additional element, using exactly the same argument:
\begin{equation*}
    f\ xs = \msf{amal}\ \left(\lambda x\to \begin{cases}f\ (\msf{filter}\ (\neq x)\ xs) & |\msf{filter}\ (\neq x) \ xs| \leq 2\\
    \msf{amal}\ (\lambda y\to f \left(\msf{filter}\ (\ne y)\ (\msf{filter}\ (\ne x)\ xs)\right) ) & \mathrm{otherwise}\end{cases}\right)
\end{equation*}
This is already sufficient to prove our Theorem -- as we will continue to stack nested \msf{filter} and \msf{amal} calls until we have only lists of no more than two unique elements, from which we can reconstruct everything we need. However, it's arguably not very satisfying, as it requires multiple nested calls to \msf{amal}.

Conveniently, we can prove a lemma that implies it is sufficient to just amalgamate once:
\begin{lemma}
Let $f : [X]\to[X]$ be \msf{filter} equivariant, and $xs : [X]$ be a list with at least four unique elements. Then it is true that:
    \[\msf{amal}\ (\lambda x\to \msf{amal}\ (\lambda y\to f\ (\msf{filter}\ (\ne \{x,y\})\ xs))) = \msf{amal}\ (\lambda (x,y)\to f\ (\msf{filter}\ (\ne\{x,y\})\ xs))\]
where, in the right hand side, we set $X'=X\times X$ and $\chi' : X'\to[X]$ to represent our collection of lists, indexed over tuples $(x, y)$ where $x\neq y$.
\end{lemma}
{\bf Proof:} We have already showed that the left-hand side is equal to $f\ xs$. First, we can show that the first element of the right-hand side matches this, i.e., $x_0(\chi')=\msf{head}\ (f\ xs)$.

Whenever $\msf{head}\ (f\ xs)\neq x\wedge \msf{head}\ (f\ xs)\neq y$, $\msf{head}\ (f\ (\msf{filter}\ (\ne \{x, y\})\ xs)) = \msf{head}\ (f\ xs)$, which satisfies the constraints needed for $x_0(\chi')$. The only thing left is to show it is unique. For any other candidate head $x'\neq \msf{head}\ (f\ xs)$, it will not be the head element in all lists where it appears -- particularly, any list indexed by $(x', y')$ where $y'\neq \msf{head}\ (f\ xs)$. Therefore, it must hold that $x_0(\chi')=\msf{head}\ (f\ xs)$.

Now, we must show that applying $\delta\chi'$ allows us to continue amalgamating in a way that gradually reconstructs $f\ xs$. Note that, because $\chi'$ uses a slightly modified input space, we need to redefine how $\delta$ applies here, by simply extending the check of which lists contain $x_0(\chi')$:
\begin{equation*}
    \delta\chi'\ (x,y) =\begin{cases}
        \chi'\ (x,y) & x=x_0(\chi')\vee y=x_0(\chi')\\
        \msf{tail}\ (\chi'\ (x,y)) & x\neq x_0(\chi')\wedge y\neq x_0(\chi')
    \end{cases}
\end{equation*}
Now, we can show that this collection is equivalent to the collections we would get by processing the tail of our desired list with the same two-element filtering, i.e.: 
\begin{equation*}\delta\chi'\ (x,y)=\msf{filter}\ (\ne \{x, y\})\ (\msf{tail}\ (f\ xs))\end{equation*}
To see this, let us examine both cases: when $x= x_0(\chi')\vee y=x_0(\chi')$, then we return $\chi'\ (x, y) = \msf{filter}\ (\neq\{x, y\})\ (f\ xs) = \msf{filter}\ (\neq\{x,y\})\ \msf{tail}\ (f\ xs)$, since the head element of $(f\ xs)$ will be filtered out in this case. When $x\ne x_0(\chi')\wedge y\ne x_0(\chi')$, then we return $\msf{tail}\ (\chi'\ (x,y)) = \msf{tail}\ (\msf{filter}\ (\ne\{x,y\})\ (f\ xs))=\msf{filter}\ (\ne\{x,y\})\ (\msf{tail}\ (f\ xs))$, where the final swap is possible because we know the first element of $f\ xs$ will not be filtered out by the \msf{filter} call.

Now, it holds (by induction on $f\ xs$, as in Lemma \ref{lem:iso}) that $\msf{amal}\ \delta\chi'=\msf{tail}\ (f\ xs)$, and we are done. 

\subsection{Amalgamation for FEs}

With the above result, it now suffices to generate all sublists of two unique elements, compute $f$ on those sublists, and then execute amalgamation to reconstruct $f\ xs$. This procedure is outlined in Figure \ref{fig:amal}, and represented by the following pseudocode:

\begin{algorithmic}

\Require \msf{filter}-equivariant $f : [X]\to[X]$
\Require $xs : X$
\Require $|X|\geq3$
\State $\chi \gets []$
\For{$X'\subset X$ s.t. $|X'| = |X|-2$}
    \State $\chi\gets (f\ (\msf{filter}\ (\notin X')\ xs)) :: \chi$
\EndFor
\State $ys\gets []$
\While{$\exists l\in\chi.\ \msf{len}\ l > 0$}
    \For{$x : X$}
        \State $\mathrm{score}(x)\gets 0$
    \EndFor
    \For{$l \in \chi$}
        \If{$\msf{len}\ l > 0$}
            \State $\mathrm{score}(\msf{head}\ l)\gets\mathrm{score}(\msf{head}\ l) + 1$
        \EndIf
    \EndFor
    \State $x_0\gets\arg\max_{x : X} \msf{score}(x)$
    \State $ys\gets x_0 :: ys$
    \For{$l \in\chi$}
        \If{$\msf{len}\ l > 0$}
            \If{$\msf{head}\ l=x_0$}
                \State $l\gets\msf{tail}\ l$
            \EndIf
        \EndIf
    \EndFor
\EndWhile
\State \Return $\msf{reverse}\ ys$
\end{algorithmic}

\subsection{Amalgamation for NFEs}
Our arguments above show that any FE function output $f\ xs$ is determined by $\{f\ ys\}$, where $ys$ ranges over all sublists of $xs$ with two unique elements. To specialise this to NFEs, first recall from Section \ref{subsec:nfes} that we can assume an input list to an NFE has unique elements -- if it doesn't already, we can uniquify with \msf{enumerate} without changing $f$'s behavior. Thus, this section's ``sublists with two unique elements'' become \emph{precisely} the sublists of length two. 

Further, given a length-two list $[x, y]$ with $x \neq y$, for any other length-two list $[p, q]$, there exists some function $g : X\rightarrow X$ with $[p, q] = \msf{map}\ g \hspace{5pt} [x, y]$. Then, \msf{map} equivariance implies that $f\ [p, q] = (\msf{map}\ g)\ (f\ [x, y])$, meaning that if we know a \emph{single} length-two example $f\ [x, y]$ it  suffices to determine \emph{every other} length-two example, and hence to determine $f$'s global behavior by amalgamation. So, we see that the ``trick'' by which an NFE like \msf{reverse} is determined from observing only one doubleton example is, in fact, mathematically sound. 

A brief counterexample shows how analogous reasoning fails for general FEs. Consider a function $f : [a] \to [a]$ that, for each unique element $x \in xs$ postpends $|x|^2$ copies of $x$ to an initially-empty output list, where $|x|$ is the number of times $x$ appears in $xs$. For example, 
\[
f \hspace{5pt} [4, 7, 4, 7, 8] = [4, 4, 4, 4, 7, 7, 7, 7, 8]
\]
We can see that $f$ is not \msf{map} equivariant because, for example, it does not commute with mapping a constant function. Further, we can see that $f$ acts as the \msf{identity} function on \emph{any} two-element list with unique elements, and hence its general behavior cannot be deduced by such examples. 

It is interesting to consider whether stronger conditions exist that rule out functions where output multiplicity depends on input multiplicity. Future work could investigate whether there are, for example, variants of $\msf{filter}\ \phi$ that, rather than retaining or discarding \emph{all} elements that match a predicate $\phi$, instead retain or discard certain subsets, or different subsets in turn. It is possible that a function which is equivariant to such functions would have the desired properties, and hence support generalisation from length-two lists.

\section{Conclusions and Further Work}

% Filter-invariant functions are interesting, eg they contain sorting functions which are known not to be map-invariant. They have a deep connection with semi-simplicial sets and their action on large inputs can be learned from their action on small inputs. The also have good structural properties, eg there are simple criteria which enable us to determine if a recursive function is filter-invariant.  

% In future work we wish to explore the connection between invariances and semi-simplicial sets, eg what is so special about lists and filter. After all, other data structures form semi-simplicial sets so have a notion of filter. We also think the amalgamation property is of fundamental interest and wonder what is the right abstract formulation of amalgamation ... we consider a sheaf theoretic formalisation will be the right way forward. And we would like to push the machine learning perspective ... what sort of other functions could we learn on small inputs and extrapolate to larger inputs.

We have presented a class of list functions for which \emph{symmetry} licenses \emph{extrapolation}. 

By the nature of extrapolation, this amounts to an account of how outputs on large examples are fully determined by knowledge of smaller ones. As another perspective on the (N)FE function classes, we can contrast them with \emph{structural recursion}, which is the more familiar way by which list functions are determined by sublists. 

Consider the definition:

\begin{Verbatim}[commandchars=\\\{\}]
\PYG{n+nf}{reverse}\PYG{+w}{ }\PYG{p}{[}\PYG{n}{x}\PYG{p}{]}\PYG{+w}{ }\PYG{o+ow}{=}\PYG{+w}{ }\PYG{p}{[}\PYG{n}{x}\PYG{p}{]}
\PYG{n+nf}{reverse}\PYG{+w}{ }\PYG{n}{x}\PYG{k+kt}{:}\PYG{n}{xs}\PYG{+w}{ }\PYG{o+ow}{=}\PYG{+w}{ }\PYG{n}{xs}\PYG{+w}{ }\PYG{o}{++}\PYG{+w}{ }\PYG{p}{[}\PYG{n}{x}\PYG{p}{]}
\end{Verbatim}

Here, it seems, we have another way of defining a function by its behavior on sublists, in this case singletons. Observe, though, that the recursive definition has \emph{two} ``degrees of freedom'': the \emph{base case}, which contains the sublist behavior, and the \emph{recursive step} function. Another function -- such as \msf{sort} -- might have the same base case, but a different recursive step. Crucially, the additional freedom in defining a step function means that recursion does not give the same length-generalisation properties as FEs: the step function itself could have pathological dependence on input length. 

This, then, offers us another way to think about FEs: they are a family of functions that share the same amalgamation function -- the ``recursive step''. Hence they leave only the degree of freedom corresponding to the base case, which in this case is behavior on lists of two unique elements.  Thus, \msf{filter} equivariance is a rigid enough condition to reduce all flexibility to the base case, but is flexible enough to still include a diverse set of interesting functions (\msf{reverse}, \msf{sort}, and so on). In future work, we can ask whether other, similarly useful constraints exist. Is there any precise sense in which FEs are the ``largest'' or most diverse class of base-case-determined functions?

We have argued that FEs support length-independent extrapolation in a quantitative sense. But it's clear that there exists functions that do not have pathological dependence on length, but which are not FEs -- some of which we discussed briefly below. Future work should give a careful account of them.

We can account for some further examples within our current framework. For example, the functions \msf{swapPairs}, which exchanges adjacent elements in a list, e.g.:
\begin{equation*}
    \msf{swapPairs}\ [1, 2, 3, 4] = [2, 1, 4, 3],
\end{equation*}
as well as \msf{swapBlocks}, which exchanges the first and second halves of a list, e.g.:
\begin{equation*}
    \msf{swapBlocks}\ [1, 2, 3, 4] = [3, 4, 1, 2],
\end{equation*}
are not \msf{filter} equivariant, but there is a precise sense in which they are \emph{products} of FE functions (\msf{reverse} and \msf{identity} in either order).

\emph{Ripple-carry addition} is another interesting example that clearly does not fit our FE framework, in particular because its local behavior is not independent across subproblems: adjacent digits have to communicate via carries. Still, this communication is itself length-independent, and we are hopeful some ``higher-order'' version of our framework could accommodate such cases. 

Just like FEs are determined by their behavior on lists of two unique elements, presumably there's a larger class of functions with their behaviour determined by actions on lists of three unique elements, and more generally, by sublists of $n$ unique elements. Can we characterise such functions, for all $n$?

Future work could also expand on the general theme of symmetries of list functions, looking beyond \msf{map} and \msf{filter}. To give some examples:
\begin{itemize}
    \item The uniquification function (\msf{nub} in Haskell) has a symmetry class different from \msf{filter} and \msf{map}. 
    \item The $\msf{double} = \msf{identity}\mdoubleplus\msf{identity}$ function has interesting symmetry class that includes some FEs, like \msf{reverse}, \msf{double} itself, and \msf{filter}, but excludes other interesting functions, like \msf{sort}.
    \item We conjecture that there is a class of functions that operates on their inputs as an undifferentiated block -- not identifying elements by their values or positions -- including \msf{double}, \msf{identity} and \msf{reverse}. We conjecture that such functions constitute the symmetry class of some collection of operators.
\end{itemize}
% While our framework goes beyond geometric deep learning \cite{bronstein2021geometric}, the structural form of the regularities we identified is very similar to equivariance. Accordingly, 

Like \msf{filter} equivariance, the transformations listed above are ``symmetries'' in the looser, monoid sense that we've discussed in this paper. But this caveat aside, one could imagine a kind of ``Erlangen program for list functions'' that would enumerate these and other symmetries, and map out how they relate to one another.

Finally, we have hinted at ways that \msf{filter} symmetry could serve as inductive bias to improve length-generalisation in machine learning models. Future work should pin this down and put it into practice.

\bibliographystyle{alpha}
\bibliography{main}

\end{document}